%% file: implicit-arap.tex
\documentclass{egpubl}
\usepackage{egsgp2026}
 
\SpecialIssuePaper         %

\CGFccby

\usepackage[T1]{fontenc}
\usepackage{dfadobe}

\usepackage{cite}  %
\BibtexOrBiblatex
\electronicVersion
\PrintedOrElectronic

\ifpdf \usepackage[pdftex]{graphicx} \pdfcompresslevel=9
\else \usepackage[dvips]{graphicx} \fi

\usepackage{egweblnk}

\input{preamble}

\title[Implicit ARAP Regularization]%
      {As-Rigid-As-Possible Regularization for Implicit Surfaces}

\author[T. Djuren, M. Worchel, U. Finnendahl \& M. Alexa]
{\parbox{\textwidth}{\centering T. Djuren$^1$\orcid{0009-0003-7043-9983}, M. Worchel$^1$\orcid{0000-0002-3469-6750}, U. Finnendahl$^1$\orcid{0000-0002-7098-1524} and M. Alexa$^1$\orcid{0000-0002-9854-8466}
        }
        \\
{\parbox{\textwidth}{\centering $^1$TU Berlin, Germany
       }
}
}

\begin{document}

\input{sections/teaser}

\maketitle

\input{sections/00_abstract}

\input{sections/01_intro}

\input{sections/02_related}

\input{sections/03_background}
\input{sections/04_method}
\input{sections/05_experiments}

\input{sections/06_discussion}

\section*{Acknowledgments}
The authors would like to thank the anonymous reviewers for their valuable comments.
This work was funded by the European Research Council (ERC) under the European Union’s Horizon 2020 research and innovation program (Grant agreement No. 101055448, ERC Advanced Grant EMERGE).

\bibliographystyle{eg-alpha-doi}  
\bibliography{bibliography}

\end{document}

%% file: preamble.tex
\usepackage{booktabs} %
\usepackage{wrapfig}
\usepackage{subcaption}
\usepackage{svg}
\usepackage{bbm}
\usepackage[most]{tcolorbox}
\usepackage{amsfonts}
\usepackage{amsmath}
\usepackage{bm}
\usepackage{xspace}
\usepackage{ccicons}

\usepackage[capitalize]{cleveref}
\crefname{algocf}{Alg.}{Algs.}
\Crefname{algocf}{Algorithm}{Algorithms}

\usepackage{csvsimple}

\usepackage{siunitx}
\usepackage{float}

\usepackage[ruled]{algorithm2e} %

\SetAlFnt{\small}
\SetAlCapFnt{\small}
\SetAlCapNameFnt{\small}
\SetAlCapHSkip{0pt}

\def\tp{^{\mathsf{T}}}
\def\Bz/{B{\'{e}}zier}
\newcommand{\arap}{\textsc{Arap}\xspace}

\newcommand{\R}{{\mathbb R}}
\newcommand{\mv}[1]{\mathbf{#1}}

\newcommand{\point}{\mv{x}}

\newcommand{\param}{\theta}

\newcommand{\loss}{\mathcal{L}}

\newcommand{\deform}{\mv{f}}
\newcommand{\deformInv}{\mv{f}^{-1}}
\newcommand{\sdf}{\Phi}

\SetKwComment{Comment}{$\triangleright\;$}{}

\DeclareCaptionFormat{custom}{
    \textbf{#1#2}\textit{#3}
}

%% file: sections/teaser.tex
\teaser{
 \includegraphics[width=\linewidth]{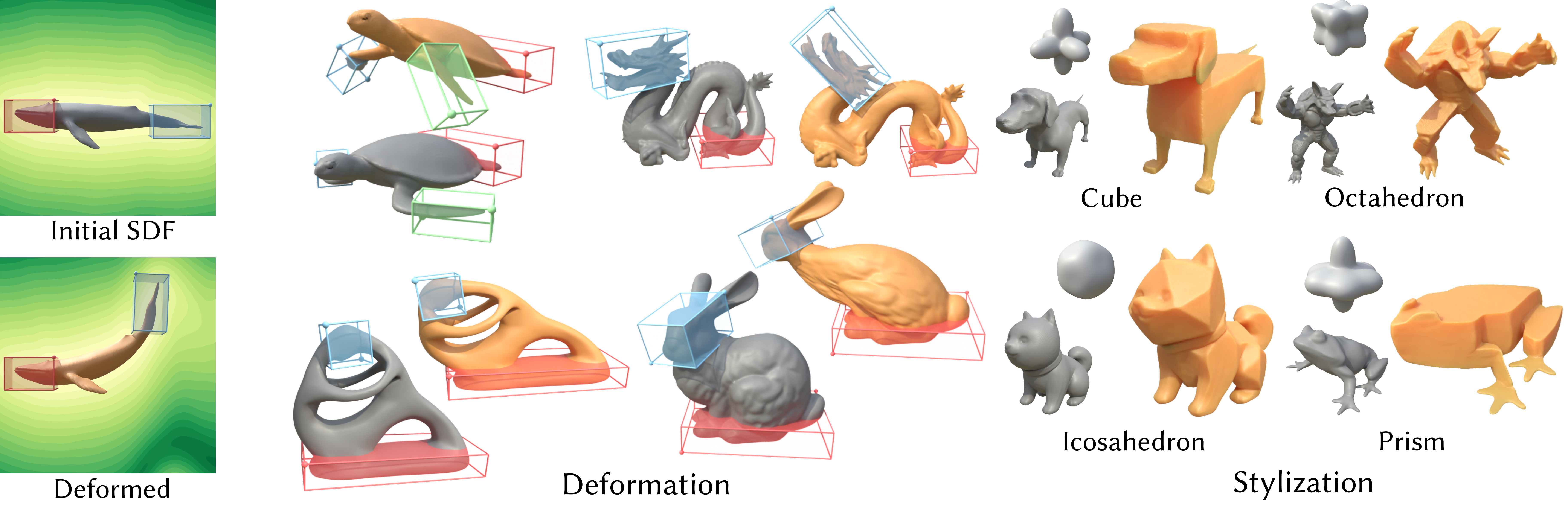}
 \centering
  \caption{We show that the well known as-rigid-as-possible (ARAP)~\cite{Sorkine2007, Chao2010} energy can be used as regularization term for implicit surface deformation by evaluating the exact point wise \arap energy using infinitesimal quantities (up to numerical precision). \arap regularization can guide the deformation of implicit surfaces e.g. using constrained regions (left) or regularize optimization procedures such as normal based stylization (right)~\cite{Liu2021, Kohlbrenner2021}.
  Blue Whale by \href{https://sketchfab.com/ostapblendercg}{Bohdan Lvov}, Hawksbill Turtle by \href{https://sketchfab.com/KNUT-OLAV_UTISTOG}{Bindestrek}, Dachshund by \href{https://sketchfab.com/PusztaiAndras}{Pusztai Andras}, Shiba Dog by \href{https://sketchfab.com/dogerlo}{zixisun02}, Frog by \href{https://sketchfab.com/Bogdan_Lapitsky}{Bogdan Lapitsky}, all under \href{https://creativecommons.org/licenses/by/4.0/}{CC-BY~\ccby}.}
\label{fig:teaser}
}

%% file: sections/00_abstract.tex
\begin{abstract}

Implicit surface representations have regained popularity because of their use in machine learning. A common component in optimization is regularization, penalizing the deviation of the surface from its original shape. The popular as-rigid-as-possible (\arap) energy strikes a good compromise between realistic deformation behavior and efficient computation, at least for piecewise linear meshes. 
We develop an approach for computing the \arap energy of a deformation function based on point sampling of the surface. The implicit representation is exploited to provide differentials in each sample. The evaluation is efficient and exact in each sample (up to numerical precision). 
We demonstrate the general applicability of the method to neural shape processing in several applications and contrast its properties with alternatives from the literature.

\ccsdesc[500]{Computing methodologies~Shape modeling}

\printccsdesc   
\end{abstract}

%% file: sections/01_intro.tex
\section{Introduction}

Many modeling tasks require regularization, penalizing the deviation of a shape from its given state. This is useful for interactive modeling, where the optimization aims to preserve the original shape while respecting user defined constraints, as well as in machine learning, where each application comes with its own optimization goals. Implicit shape representations have regained popularity in particular in the latter domain~\cite{Park2019, Mescheder2019, Sitzmann2020, Wang2021, Muller2022}, because representations map well to processing on the GPU. Baieri et al.~\cite{Baieri2025} have demonstrated that penalizing deformation based on the as-rigid as-possible (\arap) energy~\cite{Sorkine2007, Chao2010} outperforms other setups~\cite{Yang2021} in terms of both quality of the results and efficiency. Inspired by these results, we develop a computational approach that more directly transfers the principles of \arap into the realm of (neural) implicit functions. 

Our starting point is the \emph{spokes and rims} version of \arap~\cite{Chao2010}. This has two reasons: 1. For modeling with triangle meshes, this appears to be the most widely adopted version, providing consistently good results; 2. Chao et al.\ provide a second order analysis resulting in the continuous deformation energy~\cite[Sec.~4.2]{Chao2010}. We base our approach on this continuous energy and approximate it based on point samples distributed uniformly over the undeformed surface~\cite{Ling2025}. The deformation is represented as a (neural) function that needs to be differentiable. This allows us to compute the differentials for computing the energy (and its gradient for optimization) in each sample point automatically. As \arap is based on locally optimal rotations, setting this up to be accurate and efficient requires some care -- the details are described in Sec.~\ref{sec:method}.

As we show in several experiments in Sec.~\ref{sec:experiments}, our direct point sampling method based on the continuous spokes and rims \arap outperforms competing methods both in terms of quality and efficiency. It is fast enough even for quasi interactive modeling with (neural) implicits and enables other applications, where \arap serves as regularizing loss.

%% file: sections/02_related.tex
\section{Related Work}

\textit{Classical surface deformation} There is extensive research on surface deformation techniques for piecewise linear meshes, which we cannot cover here in detail~\cite{Botsch2010, Botsch2007}. Classical lattice-based freeform deformation~\cite{Sederberg1986} and cage based deformation techniques~\cite{Stroeter2024} define deformation fields in space that can be used to deform other surface representations than meshes. Usually, it is difficult with these space deformers to modify the surface directly by selecting and dragging arbitrary points of the surface. This might be one reason for the popularity of techniques that directly modify points on surfaces such as~\cite{Botsch2004} or as-rigid-as-possible (\arap) surface deformation~\cite{Sorkine2007}. Numerous works build on the \arap energy. Chao et al.~\cite{Chao2010} extend \arap with spokes-and-rims by a surface bending term and other generally improve the energy formulation~\cite{Levi2015, Oehri2025}. Recently, \arap has been adapted for Gaussian Splatting representations~\cite{Han2025}.

\textit{Neural surface deformations}  Our work is inspired by the work of Yang et al.~\cite{Yang2021}, as they implement various geometry processing tasks using a neural deformation field, including surface deformation based on differential quantities. Although they introduce terms to measure surface stretching and bending, they do not aim to minimize the well-known \arap energy. In subsequent work, Baieri et al.~\cite{Baieri2025} suggested minimizing the \arap energy for implicit surfaces instead. They learn a rotation as well as a translation field by local patch meshing the implicit surface without evaluation of differential quantities. Compared to the work by Yang et al.~\cite{Yang2021}, they claim better flexibility, robustness, and highly improved time efficiency. Another method that develops an \arap inspired energy for learning implicit shape representations with dense correspondence is~\cite{Zhang2023}. They penalize surface stretching using an equivalent term, but combine this with a hierarchy of rigid constraints. This includes sampling nearby points of the surface, which seems to have a similar effect on penalizing surface bending. One drawback to rigidly constraining nearby points is that it can impose unnecessary restrictions on other nearby parts of the surface that are not connected.

Other works predict (as-rigid-as-possible) deformations by training with datasets of deformed representations and handle positions~\cite{Tang2022, Aigerman2022}. This data based deformation could allow very fast deformations, but are restricted to deformations that where observed in the training data. There are various works on shape matching and semantic editing~\cite{Hao2020, Wang2019, Zheng2021, Deng2021, Yifan2020} that solve a matching problem, whereas we solve a matching problem with only few correspondences/handles.

%% file: sections/04_method.tex
\section{Method}
\label{sec:method}

The deformation of an implicit surface $S = \{ \point \, \vert \, \sdf(\point) = 0 \}$ defined by some function $\sdf: \R^3 \to \R$ is commonly represented using a deformation field $\deform: \R^3 \to \R^3$ that maps the \emph{initial} space to the \emph{deformed} space, or alternatively its inverse $\deform^{-1}$, which allows directly expressing the deformed implicit surface as $\tilde{S} = \{ \point \, \vert \, \sdf(\deform^{-1}(\point)) = 0 \}$ (the inverse of either function does not necessarily exist).
To make the distinction between $\deform$ and its inverse clear, we will often refer to $\deform$ as the \emph{forward} deformation function (see Fig.~\ref{fig:overview} for an overview).

Deformation fields, parameterized by $\param$ (e.g., the parameters of a neural network), are optimized in various applications in vision and graphics by minimizing a loss function
\begin{equation}
    \loss_\text{app}(\sdf, \deform_{\param})
\end{equation}
that measures some application-defined objective on the deformed surface.
It is often desirable to encourage $\deform$ to represent a `natural' deformation of the initial shape such that, for example, the deformed surface stays close to the initial surface.
Several regularization terms have been explored to achieve this goal:
some are heuristics~\cite{Selvaraju2024, Wu2024}, while others are based on established deformation energies~\cite{Yang2021}.

In particular, regularization terms based on or inspired by the well-known as-rigid-as-possible (\arap) energy~\cite{Sorkine2007} have shown promising results for implicit surfaces~\cite{Zhang2023, Baieri2025}.
Surprisingly, to the best of our knowledge, no previous work has considered the, arguably, straightforward option: an as-rigid-as-possible regularization term that directly follows from the continuous definition of the energy, including both stretching and bending.
We find that such a regularization term works surprisingly well, while, at the same time, being simple and efficient to evaluate.

In the following, we will first briefly recap the continuous as-rigid-as-possible energy as introduced by Chao et al.~\cite{Chao2010}. 
We will then derive the resulting ARAP regularization term and discuss its evaluation.

\subsection{Continuous Energy Definition}

The first continuous formulations of the as-rigid-as-possible energy can be found in the work of Chao et al.~\cite{Chao2010}.
They not only introduce the commonly used spokes-and-rims \arap but also a continuous formulation of the same energy
\begin{equation}
    E_{\text{ARAP}} = \int_S ||d \mv f-\mv{R}||_F + r^2 \langle d\mv R, d \mv R \mv Y\rangle,
    \label{eq:ARAP}
\end{equation}
defined as an integral over the undeformed surface $S$, where $||\cdot||_F$ is the Frobenius norm ($\mv R$ and $\mv Y$ will be clarified in a moment).
The energy measures intrinsic stretching and bending of the surface when it undergoes a surface deformation $\mv f: S \rightarrow \tilde{S}$.
Stretching and weighting are balanced against each other by a balancing factor $r^2 \in \R$. 
For simplicity, we discuss surfaces embedded in $\R^3$, but Eq.~\ref{eq:ARAP} also holds for curves.

\paragraph*{Stretching.}

The first term of Eq.~\eqref{eq:ARAP} penalizes stretching of the surface. The Jacobian $d\mv f = \mv J_f \in \R^{3\times 3}$ describes the change of the tangent frame at every point and $\mv R$ is the closest rotation to this transformation. The optimal closest rotation $\mv R$ to $\mv J_f$ can be computed by singular value decomposition $\mv J_f = \mv U \boldsymbol{\Sigma} \mv V^\top$ and removing the scaling and reflection contributions with replacement of $\boldsymbol{\Sigma}$:
\begin{equation}
    \mv R = \mv U \boldsymbol{\Lambda} \mv V^\top \quad \text{where} \quad \boldsymbol{\Lambda} = \text{diagonal}\big(1,1,\det(\mv U \mv V^\top)\big).
\end{equation}
The determinant $\det(\mv U \mv V^\top)$ is always $1$ or $-1$ since $\mv U$ and $\mv V$ are rotations and/or reflections. By flipping the sign of the smallest eigenvalue, if $\mv U \mv V^\top$ includes a reflection, then $\mv R$ becomes a proper rotation without reflection.

\paragraph*{Bending.}

The second term of Eq.~\eqref{eq:ARAP} penalizes surface bending by measuring the change in the closest rotations using the derivative tensor $d \mv R \in \R^{3\times3\times3}$, where the product $\langle d\mv R, d \mv R \mv Y\rangle$ can be interpreted as a sum $\sum_{\alpha} \langle d_\alpha \mv R, d_{\alpha}\mv R \mv Y\rangle$ over all dimensions $\alpha$ ($x$, $y$ and $z$). $\mv Y$ is defined by the adjusted polar decomposition $\mv J_f = \mv R \mv Y$ and can be computed using the matrices from the previous paragraph as $\mv Y = \mv V \boldsymbol{\Sigma}\boldsymbol{\Lambda} \mv V^\top$ (instead of $\mv V \boldsymbol{\Lambda} \mv V^\top$, to exclude reflection from rotation $\mv R$). Although we state an explicit bending term, the term arises implicitly from the original ARAP formulations~\cite{Chao2010}. In fact \cite{Baieri2025} minimize an approximation of the same energy implicitly as well, as their energy evaluation over patches can be interpreted as Monte Carlo integration of the energy presented by \cite{Finnendahl2026} using a box filter with radius size equal to the patch size, leading to the same bending penalty.%

\subsection{ARAP Regularization}

To turn the continuous \arap energy from Eq.~\eqref{eq:ARAP} into a regularization term, we simply use the numerical estimation of the integral.
More precisely, we use a Monte Carlo approach and estimate the integral by uniformly sampling the implicit surface, where we will denote the set of sample points on the initial implicit surface as $\mathcal{X}$. First, we will describe the regularization terms. The sampling technique is then covered in the second part of this section.

\begin{figure}
    \centering
    \includegraphics[width=\linewidth]{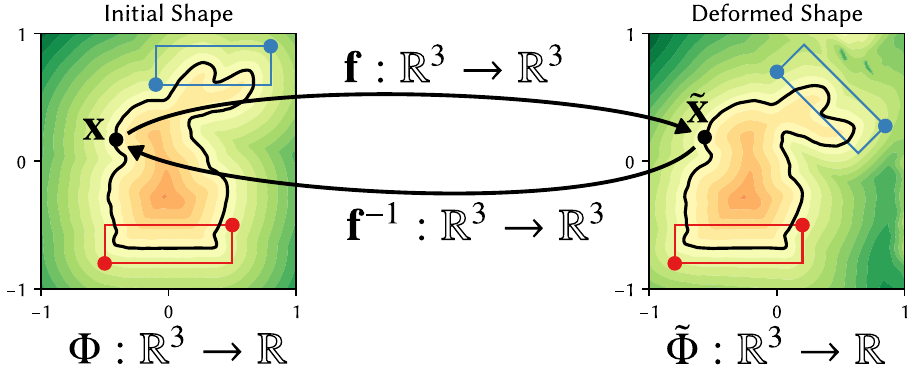}
    \caption{Given a set of constrains (colored boxes) we compute the forward deformation function $\mv f$. This allows (1) to train with samples from the initial space since the \arap energy is defined as integral over the initial surface and (2) all samples used for training can be computed before training without the need to resample points from the deformed space after each update of the deformation.}
    \label{fig:overview}
\end{figure}

At each sample point, we compute the Jacobian of the deformation $\mv J_f \in \R^{3\times 3}$. 
It can be shown that minimizing the stretching contribution as $||\mv J_f - \mv R||_F$, is equivalent to minimization of the distances of the eigenvalues of $\mv J_f$ (without reflection) to $1$~\cite{Smith2015}. This is because of the rotation invariance of the Frobenius norm. 
Thus, we directly penalize the squared stretch as
\begin{equation*}
    \loss_{\text{stretch}} = \frac{1}{|\mathcal{X}|} \sum_{\point \in \mathcal{X}} ||\boldsymbol{\Sigma}(\point)\boldsymbol{\Lambda}(\point) - \mv I||^2
\end{equation*}
$\boldsymbol{\Sigma}$ and $\boldsymbol{\Lambda}$ depend on the location sample. We highlight this as $\boldsymbol{\Sigma}(\point)$ and $\boldsymbol{\Lambda}(\point)$. To minimize the bending of the surface, we ask for vanishing derivatives of the closest rotations
\begin{equation*}
    \loss_{\text{bend}} = \frac{1}{|\mathcal{X}|} \sum_{\point \in \mathcal{X}} \langle d\mv R(\point), d \mv R(\point) \mv Y(\point)\rangle
\end{equation*}
This requires the computation of the derivatives of the rotation $\mv R$ (see below for a short discussion of the derivatives). 

\begin{figure}
    \centering
    \includegraphics[width=\linewidth]{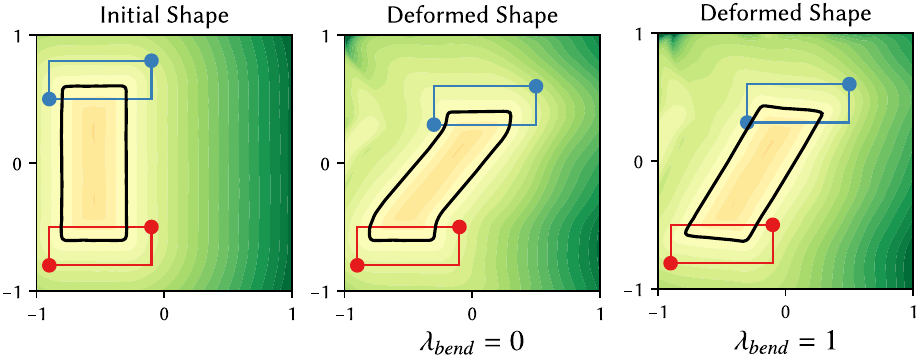}
    \caption{Regularization by the \arap energy penalizes stretching and bending of the surface/curve. The cost of bending is controlled by $\lambda_{bend}$. Constrained regions are highlighted by colored boxes and should transform to the constrained target region (same color). Deformed implicit values are generated by learning $\mv f^{-1}$ using a cycle consistency loss.}
    \label{fig:bending}
\end{figure}

The full \arap regularization is then
\begin{equation}
    \loss_{\text{ARAP}} = \loss_{\text{stretch}} + \lambda_{\text{bend}} \loss_{\text{bend}}
    \label{eq:arap_regularization}
\end{equation}
where $\lambda_{\text{bend}} = r^2 \in \R$ determines how much bending is penalized (Fig.~\ref{fig:bending}).

\paragraph*{Derivatives}

Note that we do not compute gradients of the singular value decomposition by automatic differentiation through the iterative SVD calculation for $\mv{R}$. Gradients of the $\mv U$ and $\mv V$ are not stable for small or similar singular values. In these cases, for example if initializing the network with the identity transformation, $\mv U$ and $\mv V$ are not unique and should not be used to compute gradients. We do not need gradients wrt. $\mv U$ and $\mv V$, but rather gradients wrt. the rotation $\mv R$ and use the gradient of $\mv R$ as computed in a polar decomposition instead, that is also stable for the identity transformation.

\paragraph*{Sampling}
Uniform sampling of implicit surfaces can be performed efficiently by finding all surface intersections of uniformly sampled lines in the bounding volume of the implicit surface~\cite{Ling2025, Palais2016}.
Lines can be uniformly sampled by (1) sampling a direction vector and (2) finding a support vector by sampling a point in the plane perpendicular to the direction vector such that the bounding volume is intersected by the line. 
Our sampling algorithm and implementation has two performance improvements over the original ones proposed by Ling et al.~\cite{Ling2025}. 
First, while they use rejection sampling for step (2) above, we directly sample valid lines for a given direction by uniformly sampling the (projected) faces of the bounding box. 
And second, computing all intersections of a ray with an implicit surface results in a highly unbalanced workload on GPUs when each ray is assigned to a thread. 
We perform load balancing by splitting long rays, which roughly leads to a $25\times$ speedup.

Note that the \arap energy Eq.~\ref{eq:ARAP} is originally defined as integral over the initial surface $S$. 
Therefore, the regularization term only requires samples on the initial surface $S$ and \emph{not} the deformed surface $\tilde{S}$.
We exploit this structure and precompute surface samples in the initial space in a preprocessing step.
During the optimization it is then sufficient to sample subsets from this precomputed set to avoid (potentially costly) sampling of the implicit surface in every iteration (Fig.~\ref{fig:overview}).
Although resampling from a fixed set of samples yields a biased integral estimate as some points on the surface may be missed, we found it to be sufficient for our applications.
It is also straightforward to make the estimate unbiased by injecting a few random samples in each iteration.

%% file: sections/05_experiments.tex
\section{Experiments}

\label{sec:experiments}
In the following, we demonstrate multiple use cases that benefit from regularization with the continuous \arap energy. To ensure reproducibility, the source code for the experiments, including all hyperparameter choices such as the number of samples and regularization weights, is provided in our project repository. We will start with a brief explanation of the network setup that we use throughout all experiments.

\paragraph*{Signed distance networks} We generate each signed distance function using a network $\Phi: \R^3 \rightarrow \R$ by sampling points around the surface, as well as uniform samples in the cube $[-1, 1]^3$. The networks for the implicit functions are trained by minimizing the mean square error of the predicted signed distance to the true signed distance to a fine triangle mesh.

\paragraph*{Deformation networks} Given an implicit surface function, we parameterize the deformation of $\Phi$ using the forward deformation function $\mv f: \R^3 \rightarrow \R^3$.
For any point on the initial surface $\point$ where $\sdf(\point) = 0$, the deformed surface point $\tilde{\point}$ is given by $\mv f(\point) = \tilde{\point}$.
When using a small network for $\mv f$, we can initialize $\mv f$ with the identity function such that $\mv f(\point) = \point$ and $\mv J_f = \mv I$. This is done by using a resnet like network~\cite{He2016} $\mv f(\point) = \point + \Psi(\point)$ and initializing the weights of the last layer of the network $\Psi$ with zeros. Higher order derivatives of $\mv f$ can be computed using the functional API of \textsc{Torch}~\cite{Paszke2019} or equivalent libraries.
To explicitly extract the deformed signed distance field, we suggest learning the inverse $\mv f^{-1}$ by training a second network using a cycle consistency loss 

\begin{equation*}
    \loss_{cycle} = \frac{1}{|\mathcal{Y}|} \sum_{\point \in \mathcal{Y}}{\big(\deform(\deformInv(\point)) - \point\big)^2}
\end{equation*}

\begin{figure}
    \centering
    \includegraphics[width=\linewidth]{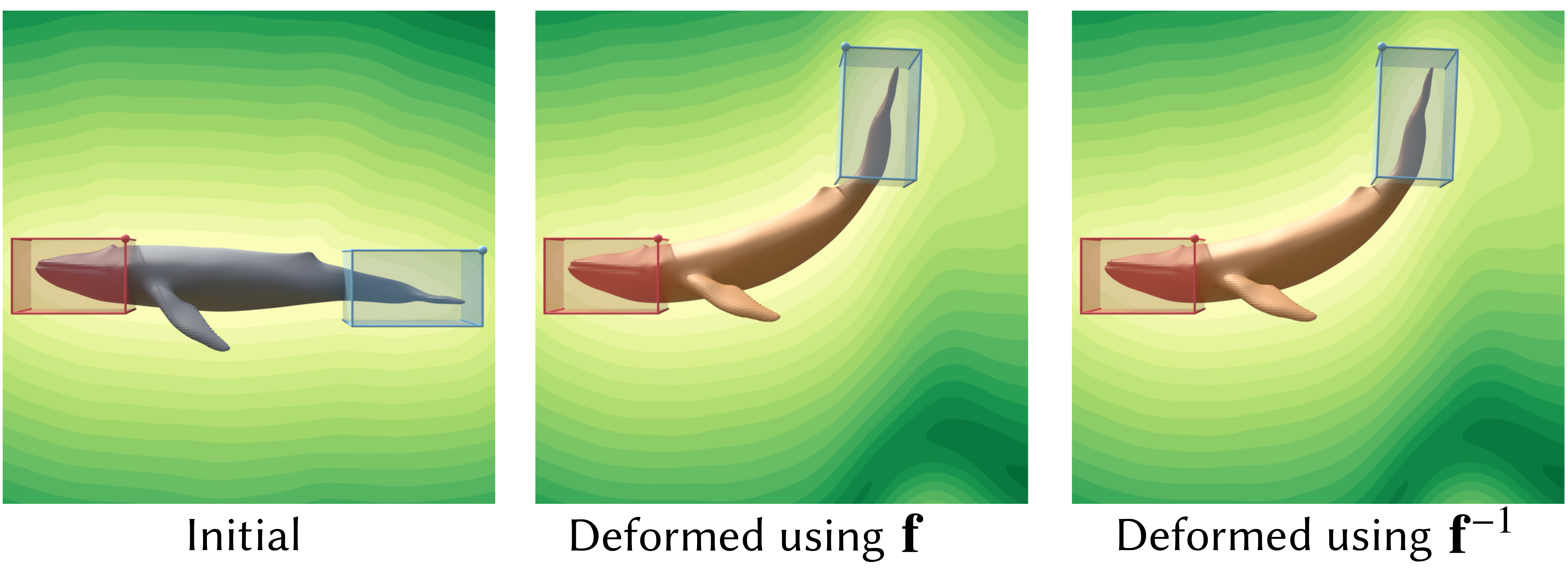}
    \caption{We compute the deformation by optimizing the forward deformation $\mv f$. To compute the deformed signed distance field, the inverse transformation is required $\tilde \Phi(\point) = \Phi(\mv f^{-1}(\point))$. We suggest to learn $\mv f^{-1}$ using a circle consistency loss. The deformed surface (middle) can be computed by marching cubes on the initial signed distance function (left) and the forward transformation $\mv f$. The result is typically indistinguishable from the marching cube mesh on $\tilde \Phi$ (right) visually. Blue Whale by \href{https://sketchfab.com/ostapblendercg}{Bohdan Lvov} under \href{https://creativecommons.org/licenses/by/4.0/}{CC-BY~\ccby}.}
    \label{fig:inverse_comparison}
\end{figure}

The set $\mathcal{Y}$ of samples used for the cycle consistency loss is the union of 1) the deformed points of $\mathcal{X}$ as well as 2) a few uniform samples in the deformed space. Additional points in the deformed space are required because the deformed space is sampled evenly for marching cubes and the deformed surface samples usually do not cover the deformed space uniformly. This encourages $\mv f$ to be invertible and prevents positions in the deformed space that are not on the surface from accidentally mapping to a point on the surface in the initial space. There are no other constraints for the inverse network and we observed that surfaces extracted using the inverse function are very close to surfaces that were extracted by the forward mapping (Fig.~\ref{fig:inverse_comparison}).

\paragraph*{Architecture} For all networks (deformation, inverse of deformation and signed distance functions), we decided to use simple fully connected networks with four hidden layers, each with $256$ neurons. We use softplus activations (with sharpness parameter $30$) because a smooth activation function is required to compute higher order derivatives.
For all networks, we use a feature encoding of the input coordinates as in~\cite{Mildenhall2021}
\begin{equation*}
    \gamma(\point) = (\sin(2^0 \pi \point), \cos(2^0 \pi \point), \dots, \sin(2^{L-1} \pi \point), \cos(2^{L-1} \pi \point))
\end{equation*}
Alternatively to learning two networks for the forward mapping $\deform$ and the $\deformInv$ mapping, one can also use invertible network architectures~\cite{Behrmann2019} as for example in~\cite{Baieri2025, Yang2021}. Note that the choice of network architecture is largely independent of \arap regularization.

As hyperparameter, we choose $L=32$ and train all networks using the \textsc{Adam} optimizer~\cite{Kingma2017} with a learning rate of $0.001$.
Our implementation will be available alongside the publication.

\paragraph*{Surface visualization} Unless otherwise noted, we visualize all deformed surfaces, by extracting the zero level set of the initial mesh using marching cubes~\cite{Lorensen1987} and deform the result by deformation $\mv f$. All surfaces in all figures were extracted using marching cubes with a resolution of $256^3$ (and possibly deformed by $\mv f$). To visualize the deformed indicator functions, we use the inverse network.

\input{sections/subsections/surface_modelling}

\input{sections/subsections/gauss_stylization}

%% file: sections/subsections/surface_modelling.tex
\subsection{Implicit Surface Modeling}

\begin{figure*}[htb]
    \centering
    \includegraphics[width=0.9\linewidth]{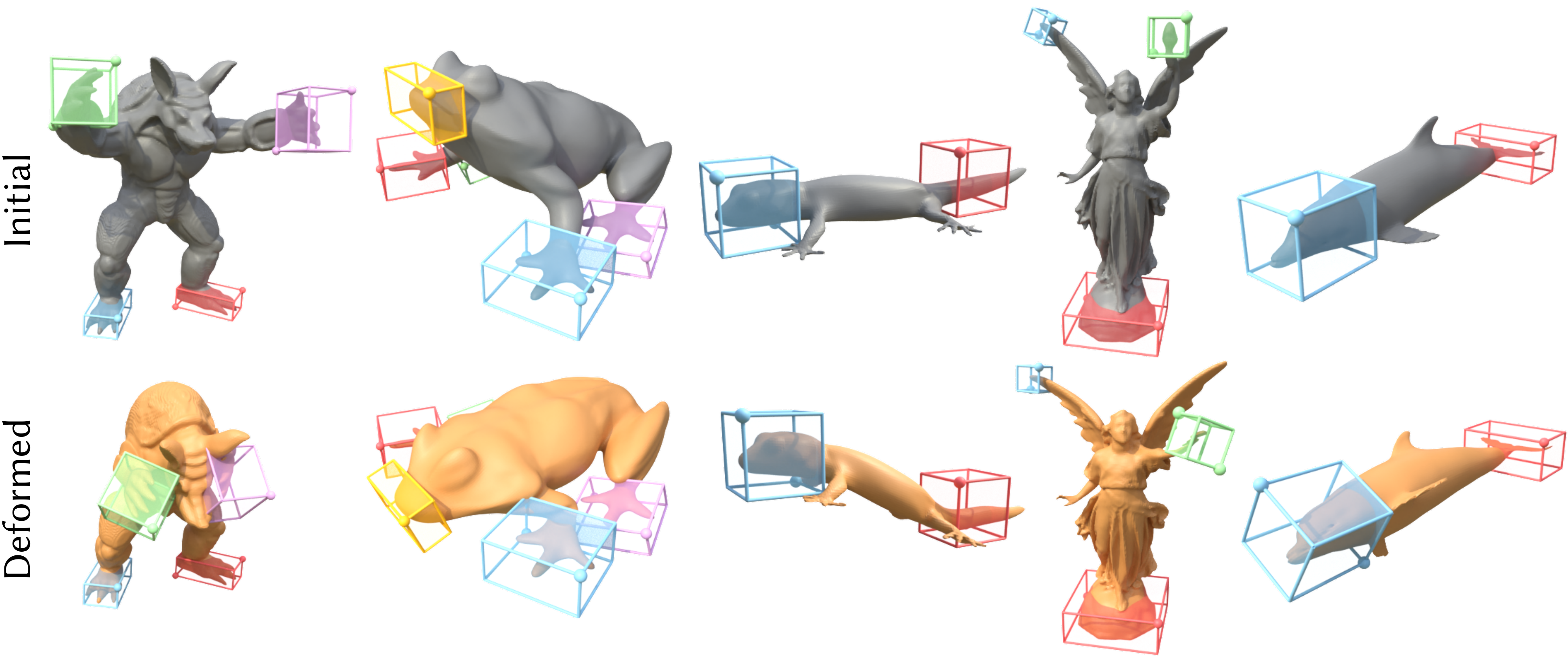}
    \caption{Implicit \arap regularization allows to model implicit functions by defining constrained regions and minimizing the as-rigid-as possible energy. The top row shows the initial implicitly defined surfaces (extracted by marching cubes~\cite{Lorensen1987}) and the deformed surfaces are in the bottom row. Whereas in original \arap for meshes~\cite{Sorkine2007} the constraints are usually defined as sets of vertices, for implicit \arap regularization we define the constraints as continuous regions (here as a set of cuboids). Each surface point in a cuboid of the initial surface space should undergo the same rigid transformation as the cuboid (same color). Frog by \href{https://sketchfab.com/Bogdan_Lapitsky}{Bogdan Lapitsky}, Tokay Gecko and Bottlenose Dolphin by \href{https://sketchfab.com/DigitalLife3D}{DigitalLife3D}, all under under \href{https://creativecommons.org/licenses/by/4.0/}{CC-BY~\ccby}.}
    \label{fig:constrained_deformation}
\end{figure*}

The classic application for \arap is surface deformation by defining sets of constraints that determine how a given surface should be deformed. 
Traditionally, constraints are a set of surface points on the initial mesh as well as their target positions on the deformed mesh.
In our setting, the surface is defined implicitly and the goal is to optimize a deformation field $\deform$ under the \arap regularization energy $\loss_{\text{ARAP}}$ and some energy $\loss_{\text{constr}}$ that represents the constraints.
In a slight generalization, we constrain surfaces within continuous regions of space, which is similar to the common approach of constraining multiple neighboring vertices in a mesh.
We define these constraint regions using a set of cuboids, where all points inside a cuboid $\mathcal{C}_i$ are constrained to deform according to a transformation $\mv C_i \in \R^{3 \times 3}$. %
The constrained position of a point $\point$ inside a cuboid is $\tilde{\point}_i =\mv C_i \, \point$, which allows us to define our constraint loss as
\begin{equation*}
    \loss_{\text{constr}} = \frac{1}{n} \sum_{i = 1}^n \frac{1}{|\mathcal{X} \cap \mathcal{C}_i|} \sum_{\point \in \mathcal{X} \cap \mathcal{C}_i} \big(\deform(\point) - \tilde{\point}_i \big)^2.
\end{equation*}
Notice that it is still possible to enforce point-wise constraints as the limit case of a constraint region that contains only a single point.
Since we precompute the set of points $\mathcal{X}$ on the initial surface, we can also precompute whether a point is constrained $\point \in \mathcal{X} \cap \mathcal{C}_i$ and its constrained position $\tilde{\point}_i$. 

We show various results for constraints and deformed implicit surfaces with different numbers and sizes of constraints, as well as different types of movement induced by constraints in Fig.~\ref{fig:constrained_deformation}. Further examples, e.g., for strong deformation like that of the bunny, are in Fig.~\ref{fig:teaser}.
Since we use a simple and relatively lightweight deformation network as well as precomputed samples, the optimization runs at interactive rates (see the supplemental video). All frames in the supplemental video are at the respective timestamps of the optimization, excluding the time taken for surface extraction using marching cubes (with resolution $256^3$) and rendering in Blender~\cite{Blender}. We used 32-bit floating-point precision on a 12th Gen Intel Core i9-12900K CPU with an Nvidia RTX 5090 GPU.

\paragraph*{Comparison to Implicit-ARAP}

\begin{figure}[htb]
    \centering
    \includegraphics[width=\linewidth]{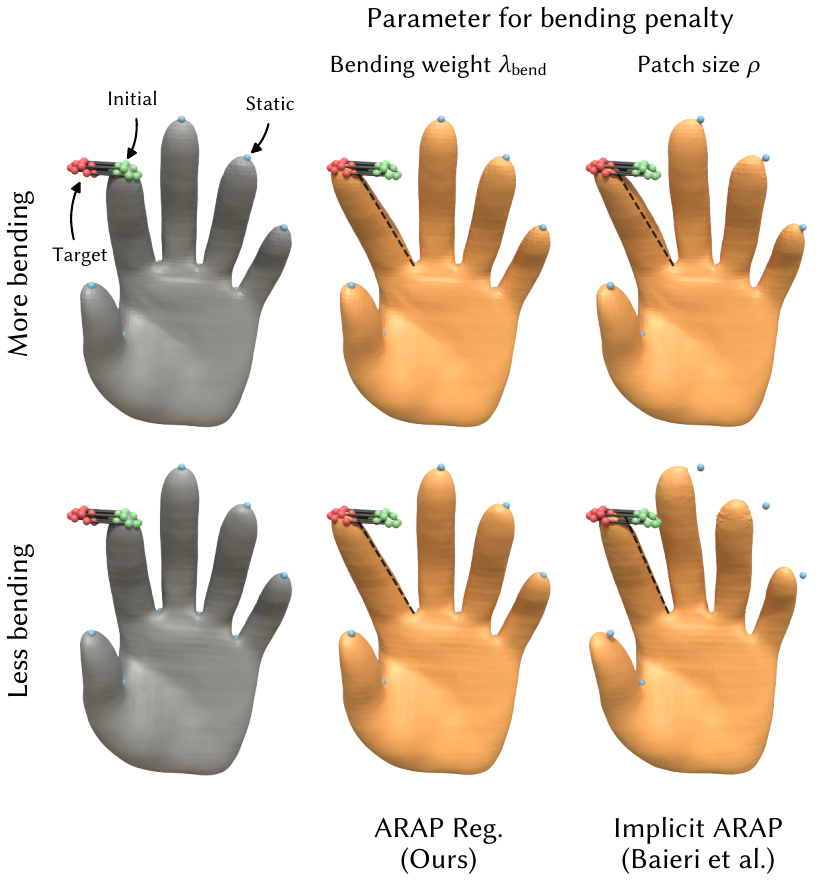}
    \caption{Implicit surface modeling with point constraints and controlled bending using different \arap{}-based approaches. Implicit-\arap{} \cite{Baieri2025} evaluates a discrete \arap{} energy by locally approximating level sets using patches (right column). 
    Since the patches only approximate the level set surfaces, topologically distant regions on the surface may be coupled and are not deformed independently (all fingers move with the index finger).
    Increasing the patch size is the only way to penalize bending, which, however, amplifies the effect.
    Our regularization term is derived from the continuous energy and only requires point samples (middle column).
    The bending weight $\lambda_\text{bend}$ can be used to directly penalize bending without the above compromise.}
    \label{fig:iarap_comparison_hand}
\end{figure}

Deforming implicit surfaces using an \arap{}-based approach for the purpose of modeling has previously been explored by Baieri et al.~\shortcite{Baieri2025} (``Implicit-\arap{}'').
The method locally meshes the implicit surface and uses the so generated patches to evaluate a \emph{discrete} \arap{} energy.
However, these patches do not strictly reflect the actual surface topology because they are generated using a simple nearest point projection from the tangent space of a sample position to the level set.
The patch size therefore becomes an intricate parameter that couples the effects of bending penalty and approximation error: very small patches effectively do not penalize bending, but the surface is well-approximated; larger patches penalize bending, yet the surface is poorly approximated. 
In the latter case, topologically distant regions are coupled such that they cannot be deformed independently, even though the deformation network would be able to separate them (Fig.~\ref{fig:iarap_comparison_hand}, right column).
For our regularization term derived from the \emph{continuous} energy, evaluating point samples suffices.
This is not only simpler as it avoids having to locally mesh the implicit surface, but it also exposes a parameter that directly controls the bending penalty (c.f. Eq.\eqref{eq:arap_regularization}), without having to balance two competing objectives (Fig.~\ref{fig:iarap_comparison_hand}, middle column).

\begin{figure}[htb]
    \centering
    \includegraphics[width=\linewidth]{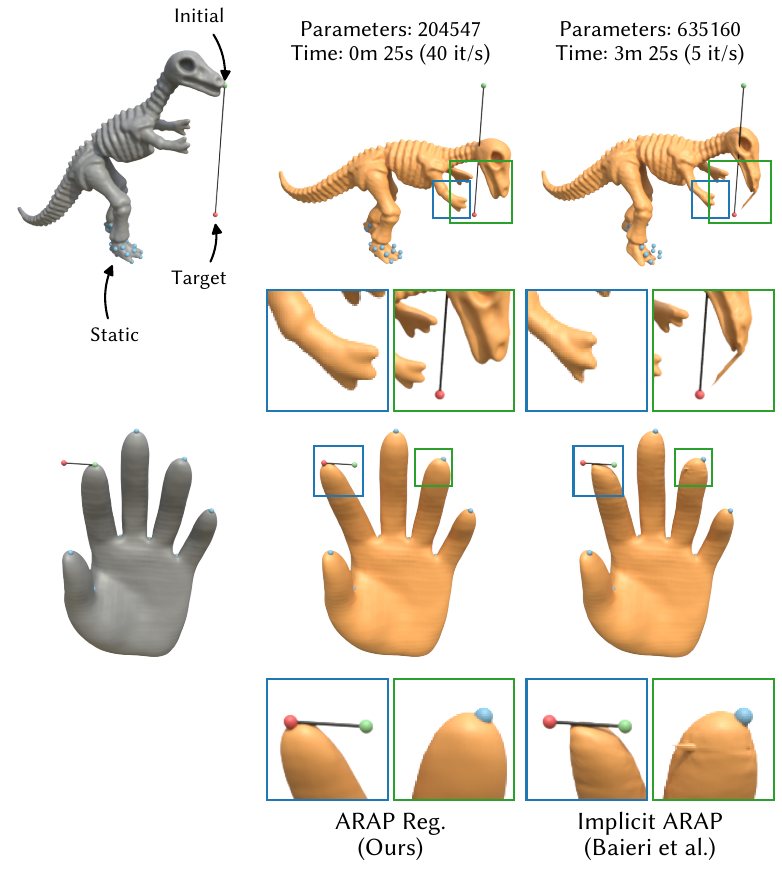}
    \caption{Implicit surface modeling with point constraints using different \arap{}-based approaches.
    The \arap{} regularization term of Implicit-\arap{} \cite{Baieri2025} (right column) is restricted to deformation fields expressed as point-wise roto-translations, while our regularization term (middle column) does not impose such restrictions.
    When the same network architecture is used for the signed distance function, our regularization results in better deformations even with a more lightweight deformation network.
    By using a simple deformation network architecture (e.g., without strict invertibility) and precomputing surface samples, we achieve interactive optimization runtime.}
    \label{fig:iarap_comparison_single}
\end{figure}

The \arap{} regularization term formulated by Baieri et al.~\shortcite{Baieri2025} is strictly restricted to deformation fields that are expressed as point-wise roto-translations: the rotation at a point serves both as part of the deformation \emph{and} as the ``optimal'' rotation used in their regularization term.
We are unsure about the potential effects of this coupling, in particular, because the predicted rotations are rotations about the origin instead of (local) rotations of the tangent frame.
In our experiments, we did not observe any benefits that would justify the resulting restrictions, rather the opposite: simply optimizing a translation for the deformation field using our regularization term leads to better deformation results, even with much smaller neural networks (Fig.~\ref{fig:iarap_comparison_single}).

%% file: sections/subsections/gauss_stylization.tex
\subsection{Implicit Gauss Stylization}

\begin{figure*}[htb]
    \centering
    \includegraphics[width=0.9\linewidth]{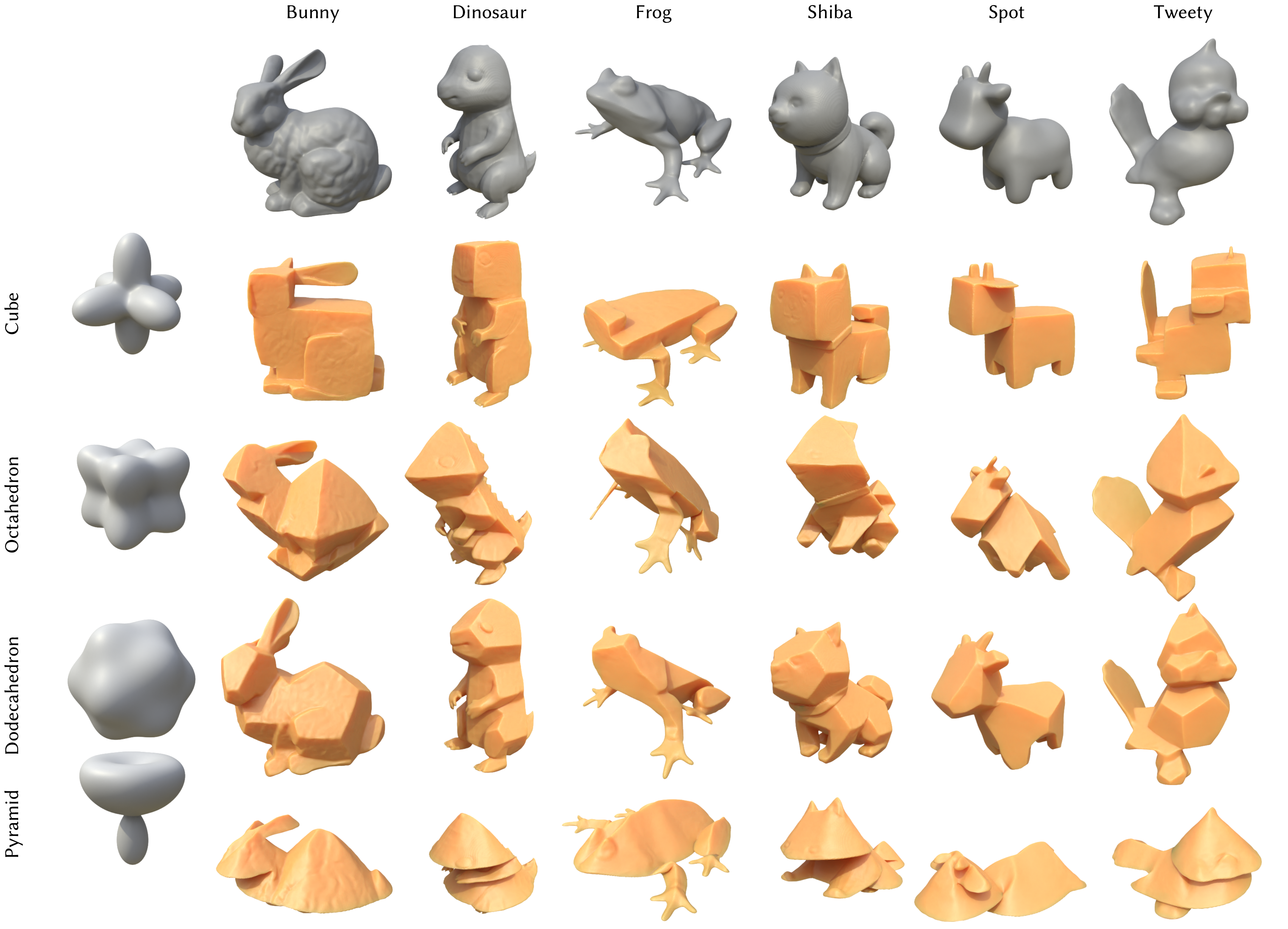}
    \caption{In analogy to Gauss Stylization~\cite{Kohlbrenner2021} and normal-driven spherical shape analogies~\cite{Liu2021}, we can use preference functions $g$ Eq.~\ref{eq:gauss_g} on the Gauss map (first column) as well as implicit \arap regularization to create diverse, continuously defined stylizations for implicit surfaces. The implicitly defined surfaces (their marching cube meshes~\cite{Lorensen1987}) that where used as input are in the first row.
    Dinosaur and Shiba Dog by \href{https://sketchfab.com/dogerlo}{zixisun02}, Frog by \href{https://sketchfab.com/Bogdan_Lapitsky}{Bogdan Lapitsky}, all under \href{https://creativecommons.org/licenses/by/4.0/}{CC-BY~\ccby}.}
    \label{fig:gauss_stylization}
\end{figure*}

There are multiple methods to stylize meshes for artistic purposes or to compute polycubes~\cite{Huang2014, Tarini2004}. One of the first of these stylization methods is Cubic stylization~\cite{Liu2019} followed by multiple generalizations to arbitrary preference functions on the Gauss map~\cite{Liu2021, Kohlbrenner2021, Binnninger2021}. All these approaches define energy functions based on the normals of the mesh and the \arap energy. In the continuous setting, we can compute normals of the implicit function by differentiation of the signed distance function $\mv n(\point) = \nabla \Phi(\point) / ||\nabla \Phi(\point)||$. These normals can again be precomputed for all sampled points before optimization. The normals after deformation with $\mv f$ are given using the inverse transpose of the Jacobian of $\mv f$, $\tilde{\mv n}(\point) = \mv J_f(\point)^{-\top} \mv n(\point)$.

We adapt the explicit preference functions from~\cite{Kohlbrenner2021} to define a continuous Gauss Stylization loss $\loss_{\text{Gauss}}$ which we can optimize for implicit surfaces using \arap regularization. For a set of $k$ preferred normal directions $\mv n_k$, we adapt the preference function of Kohlbrenner et al.~\cite{Kohlbrenner2021} that is based on the Mises-Fisher distribution as follows

\begin{equation}
    g(\mv n) = \sum_k w_k \exp(\sigma \mv n(\point)^\top \mv n_k).
    \label{eq:gauss_g}
\end{equation}

\setlength{\intextsep}{0pt}%
\setlength{\columnsep}{5pt}%
\begin{wrapfigure}{r}{0.3\linewidth}
    \includegraphics[width=\linewidth]{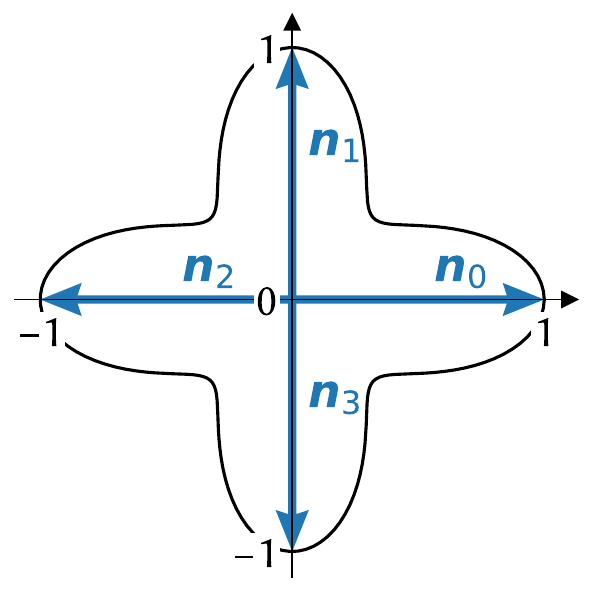}
\end{wrapfigure}

Typically, $\mv n_k$ is a set of, for example, $6$ normals of a cube or $12$ normals of a dodecahedron. $w_k$ provide control over the influence of the individual normals and $\sigma$ determines the scale of the preference function (see inset). We choose $w_k$ by solving a small linear system of equations such that the preference function is $1$ in preferred normal directions $g(\mv{n}_k) = 1$ (see~\cite{Kohlbrenner2021}):

\begin{equation}
\begin{pmatrix}
\exp(\sigma) & 
\exp (\sigma\mv{n}_0\tp \mv{n}_1) &
\exp (\sigma\mv{n}_0\tp \mv{n}_2) &
\hdots\\
\exp (\sigma\mv{n}_1\tp \mv{n}_0) &
\exp(\sigma) & 
\exp (\sigma\mv{n}_1\tp \mv{n}_2) &
\hdots\\
\vdots & \vdots & \vdots & \ddots\\
\end{pmatrix}
\mv w
=
\mv 1
\end{equation}

During the training procedure, we evaluate $g$ at the sample points and define the Gauss stylization loss as

\begin{equation*}
    \loss_{\text{Gauss}} = \frac{1}{|\mathcal{X}|} \sum_{\point \in \mathcal{X}} g\left(\mv J_f(\point)^{-\top} \frac{\nabla \Phi(\point)}{||\nabla \Phi(\point)||}\right)
\end{equation*}

Minimizing the Gauss Stylization loss together with \arap regularization determines the shape up to translation. The global rotation of the optimized shape is determined by the Gauss loss.  To keep the shape in place during optimization, we choose one sample point $\point_{\text{fix}}$ at the bottom of the implicit surface, which should stay fixed using a third loss term.

\begin{equation*}
    \loss_{\text{fix}} = ||\mv f(\point_{\text{fix}}) - \point_{\text{fix}}||^2
\end{equation*}

The final loss for implicit Gauss Stylization is then:

\begin{equation*}
    \loss = \lambda_{\text{Gauss}} \ \loss_{\text{Gauss}} + \lambda_{\text{ARAP}} \ \loss_{\text{ARAP}} + \loss_{\text{fix}}
\end{equation*}

Similarly to $\lambda_{\text{Gauss}}$, the parameter $\sigma$ determines the scale of the normal preference function. We fix $\sigma=5$ to keep the regularization parameters $\lambda$ in a similar range. We demonstrate implicit Gauss Stylization for various implicit shapes and preferred normal directions in Fig.~\ref{fig:gauss_stylization} and Fig.~\ref{fig:teaser}.

%% file: sections/06_discussion.tex
\section{Discussion}

We think that regularization with the \arap energy is a valuable tool in geometry processing for implicit surfaces that include measuring the deformation between surfaces. It seems reasonable to derive the deformation energy for stretching and bending from known continuous \arap formulations~\cite{Sorkine2007, Chao2010} that are widely used for mesh deformation. All relevant computations can be performed directly using differential quantities of the forward mapping. These quantities can be computed without local patch meshing or additional resampling around each sample point, which is how previous work evaluated \arap inspired energies. This eliminates artifacts caused by poorly fitted local triangle patches for complex shapes and simplifies the energy computation. Local patches also lead to piecewise linear approximation of the energy at each sample point, whereas this method computes the exact \arap energy (up to numerical precision). At the same time, it is possible to achieve almost interactive processing times.
We demonstrate \arap regularization for a variety of deformations and meshes, and we show exemplarily how this can guide optimization in a surface stylization setting.

Limitations are that the constraints are only approximated, whereas in the mesh based \arap energy constraints are interpolated and this can only be partially prevented by assigning a higher weight the interpolation loss.The missing interpolation property can be particularly apparent when a penalized intermediate state of self intersection prevents the deformation of the surface. With the cycle consistency loss we penalize multiple points mapping to the same location, whereas multiple deformed vertices and constraints can occupy the same position in \arap~\cite{Sorkine2007}. Continuous \arap regularization might produce to unexpected results when multiple constraints regions overlap. In addition, we think that the computation of gradients can be accelerated by explicit gradient computations that avoid computing the entire singular decomposition. The computation of all orthogonal matrices is not required when the rotation is obtained by polar decomposition. Although we did not observe artifacts caused by computing higher-order derivatives, it has been shown that higher derivatives computed by automatic differentiation can include noise, and ways to improve this are presented in~\cite{Chetan2025}.
In this work, we focus on the application of the \arap energy to implicit surfaces as those representations become increasingly popular because of their use in machine learning. However, this is largely independent of the representation of the surface, and deforming other representations such as parametric surfaces appears to be straightforward.

%% file: bibliography.bib
@misc{Baieri2025,
title={Implicit-ARAP: Efficient Handle-Guided Neural Field Deformation via Local Patch Meshing}, 
author={Daniele Baieri and Filippo Maggioli and Emanuele Rodolà and Simone Melzi and Zorah Lähner},
year={2025},
NOeprint={2405.12895},
archivePrefix={arXiv},
primaryClass={cs.GR},
NOurl={https://arxiv.org/abs/2405.12895}, 
}

@inproceedings{Yang2021,
 author = {Yang, Guandao and Belongie, Serge and Hariharan, Bharath and Koltun, Vladlen},
 booktitle = {Advances in Neural Information Processing Systems},
 editor = {M. Ranzato and A. Beygelzimer and Y. Dauphin and P.S. Liang and J. Wortman Vaughan},
 pages = {22483--22497},
 publisher = {Curran Associates, Inc.},
 title = {Geometry Processing with Neural Fields},
 NOurl = {https://proceedings.neurips.cc/paper_files/paper/2021/file/bd686fd640be98efaae0091fa301e613-Paper.pdf},
 volume = {34},
 year = {2021}
}

@article{Chao2010,
author = {Chao, Isaac and Pinkall, Ulrich and Sanan, Patrick and Schr\"{o}der, Peter},
title = {A simple geometric model for elastic deformations},
year = {2010},
issue_date = {July 2010},
publisher = {Association for Computing Machinery},
address = {New York, NY, USA},
volume = {29},
number = {4},
issn = {0730-0301},
NOurl = {https://doi.org/10.1145/1778765.1778775},
doi = {10.1145/1778765.1778775},
journal = {ACM Trans. Graph.},
month = jul,
articleno = {38},
numpages = {6}
}

@inproceedings{Sorkine2007,
title={As-rigid-as-possible surface modeling},
author={Sorkine, Olga and Alexa, Marc and others},
booktitle={Symposium on Geometry processing},
volume={4},
pages={109--116},
year={2007}
}

@ARTICLE{Levi2015,
author={Levi, Zohar and Gotsman, Craig},
journal={IEEE Transactions on Visualization and Computer Graphics}, 
title={Smooth Rotation Enhanced As-Rigid-As-Possible Mesh Animation}, 
year={2015},
volume={21},
number={2},
pages={264-277},
doi={10.1109/TVCG.2014.2359463}
}

@incollection{Paszke2019,
title = {PyTorch: An Imperative Style, High-Performance Deep Learning Library},
author = {Paszke, Adam and Gross, Sam and Massa, Francisco and Lerer, Adam and Bradbury, James and Chanan, Gregory and Killeen, Trevor and Lin, Zeming and Gimelshein, Natalia and Antiga, Luca and Desmaison, Alban and Kopf, Andreas and Yang, Edward and DeVito, Zachary and Raison, Martin and Tejani, Alykhan and Chilamkurthy, Sasank and Steiner, Benoit and Fang, Lu and Bai, Junjie and Chintala, Soumith},
booktitle = {Advances in Neural Information Processing Systems 32},
pages = {8024--8035},
year = {2019},
publisher = {Curran Associates, Inc.},
NONOurl = {http://papers.neurips.cc/paper/9015-pytorch-an-imperative-style-high-performance-deep-learning-library.pdf}
}

@article{Mildenhall2021,
author = {Mildenhall, Ben and Srinivasan, Pratul P. and Tancik, Matthew and Barron, Jonathan T. and Ramamoorthi, Ravi and Ng, Ren},
title = {NeRF: representing scenes as neural radiance fields for view synthesis},
year = {2021},
issue_date = {January 2022},
publisher = {Association for Computing Machinery},
address = {New York, NY, USA},
volume = {65},
number = {1},
issn = {0001-0782},
NOurl = {https://doi.org/10.1145/3503250},
doi = {10.1145/3503250},
journal = {Commun. ACM},
month = dec,
pages = {99–106},
numpages = {8}
}

@misc{Kingma2017,
title={Adam: A Method for Stochastic Optimization}, 
author={Diederik P. Kingma and Jimmy Ba},
year={2017},
Noeprint={1412.6980},
archivePrefix={arXiv},
primaryClass={cs.LG}
}

@article{Ling2025,
author = {Ling, Selena and Madan, Abhishek and Sharp, Nicholas and Jacobson, Alec},
title = {Uniform Sampling of Surfaces by Casting Rays},
journal = {Computer Graphics Forum},
volume = {44},
number = {5},
pages = {e70202},
doi = {https://doi.org/10.1111/cgf.70202},
NOurl = {https://onlinelibrary.wiley.com/doi/abs/10.1111/cgf.70202},
Noeprint = {https://onlinelibrary.wiley.com/doi/pdf/10.1111/cgf.70202},
year = {2025}
}

@misc{Palais2016,
title={PointClouds: Distributing Points Uniformly on a Surface}, 
author={Richard Palais and Bob Palais and Hermann Karcher},
year={2016},
Noeprint={1611.04690},
archivePrefix={arXiv},
primaryClass={math.DG},
NOurl={https://arxiv.org/abs/1611.04690}, 
}

@InProceedings{He2016,
author = {He, Kaiming and Zhang, Xiangyu and Ren, Shaoqing and Sun, Jian},
title = {Deep Residual Learning for Image Recognition},
booktitle = {Proceedings of the IEEE Conference on Computer Vision and Pattern Recognition (CVPR)},
month = {June},
year = {2016}
}

@article{Liu2019,
author = {liu, Hsueh-Ti Derek and Jacobson, Alec},
title = {Cubic stylization},
year = {2019},
issue_date = {December 2019},
publisher = {Association for Computing Machinery},
address = {New York, NY, USA},
volume = {38},
number = {6},
issn = {0730-0301},
NOurl = {https://doi.org/10.1145/3355089.3356495},
doi = {10.1145/3355089.3356495},
journal = {ACM Trans. Graph.},
month = nov,
articleno = {197},
numpages = {10}
}

@article{Liu2021,
author = {Liu, Hsueh-Ti Derek and Jacobson, Alec},
title = {Normal-Driven Spherical Shape Analogies},
journal = {Computer Graphics Forum},
volume = {40},
number = {5},
pages = {45-55},
doi = {https://doi.org/10.1111/cgf.14356},
NOurl = {https://onlinelibrary.wiley.com/doi/abs/10.1111/cgf.14356},
Noeprint = {https://onlinelibrary.wiley.com/doi/pdf/10.1111/cgf.14356},
year = {2021}
}

@article{Kohlbrenner2021,
author = {Kohlbrenner, M. and Finnendahl, U. and Djuren, T. and Alexa, M.},
title = {Gauss Stylization: Interactive Artistic Mesh Modeling based on Preferred Surface Normals},
journal = {Computer Graphics Forum},
volume = {40},
number = {5},
pages = {33-43},
doi = {https://doi.org/10.1111/cgf.14355},
NOurl = {https://onlinelibrary.wiley.com/doi/abs/10.1111/cgf.14355},
Noeprint = {https://onlinelibrary.wiley.com/doi/pdf/10.1111/cgf.14355},
year = {2021}
}

@INPROCEEDINGS {Selvaraju2024,
author = { Selvaraju, Pratheba },
booktitle = { 2024 International Conference on 3D Vision (3DV) },
title = {{ Developability Approximation for Neural Implicits Through Rank Minimization }},
year = {2024},
volume = {},
ISSN = {},
pages = {780-789},
doi = {10.1109/3DV62453.2024.00041},
NOurl = {https://doi.ieeecomputersociety.org/10.1109/3DV62453.2024.00041},
publisher = {IEEE Computer Society},
address = {Los Alamitos, CA, USA},
month =mar}

@article{Binnninger2021,
author = {Binninger, Alexandre and Verhoeven, Floor and Herholz, Philipp and Sorkine-Hornung, Olga},
title = {Developable Approximation via Gauss Image Thinning},
journal = {Computer Graphics Forum},
volume = {40},
number = {5},
pages = {289-300},
doi = {https://doi.org/10.1111/cgf.14374},
NOurl = {https://onlinelibrary.wiley.com/doi/abs/10.1111/cgf.14374},
Noeprint = {https://onlinelibrary.wiley.com/doi/pdf/10.1111/cgf.14374},
year = {2021}
}

@inproceedings{Lorensen1987,
author = {Lorensen, William E. and Cline, Harvey E.},
title = {Marching cubes: A high resolution 3D surface construction algorithm},
year = {1987},
isbn = {0897912276},
publisher = {Association for Computing Machinery},
address = {New York, NY, USA},
NOurl = {https://doi.org/10.1145/37401.37422},
doi = {10.1145/37401.37422},
booktitle = {Proceedings of the 14th Annual Conference on Computer Graphics and Interactive Techniques},
pages = {163–169},
numpages = {7},
series = {SIGGRAPH '87}
}

@book{Botsch2010,
  title={Polygon mesh processing},
  author={Botsch, Mario and Kobbelt, Leif and Pauly, Mark and Alliez, Pierre and L{\'e}vy, Bruno},
  year={2010},
  publisher={CRC press}
}

@article{Botsch2007,
  title={On linear variational surface deformation methods},
  author={Botsch, Mario and Sorkine, Olga},
  journal={IEEE transactions on visualization and computer graphics},
  volume={14},
  number={1},
  pages={213--230},
  year={2007},
  publisher={IEEE}
}

@inproceedings{Sederberg1986,
author = {Sederberg, Thomas W. and Parry, Scott R.},
title = {Free-form deformation of solid geometric models},
year = {1986},
isbn = {0897911962},
publisher = {Association for Computing Machinery},
address = {New York, NY, USA},
NOurl = {https://doi.org/10.1145/15922.15903},
doi = {10.1145/15922.15903},
booktitle = {Proceedings of the 13th Annual Conference on Computer Graphics and Interactive Techniques},
pages = {151–160},
numpages = {10},
series = {SIGGRAPH '86}
}

@article{Stroeter2024,
author = {Ströter, D. and Thiery, J. M. and Hormann, K. and Chen, J. and Chang, Q. and Besler, S. and Mueller-Roemer, J. S. and Boubekeur, T. and Stork, A. and Fellner, D. W.},
title = {A Survey on Cage-based Deformation of 3D Models},
journal = {Computer Graphics Forum},
volume = {43},
number = {2},
pages = {e15060},
doi = {https://doi.org/10.1111/cgf.15060},
NOurl = {https://onlinelibrary.wiley.com/doi/abs/10.1111/cgf.15060},
Noeprint = {https://onlinelibrary.wiley.com/doi/pdf/10.1111/cgf.15060},
year = {2024}
}

@article{Botsch2004,
author = {Botsch, Mario and Kobbelt, Leif},
title = {An intuitive framework for real-time freeform modeling},
year = {2004},
issue_date = {August 2004},
publisher = {Association for Computing Machinery},
address = {New York, NY, USA},
volume = {23},
number = {3},
issn = {0730-0301},
NOurl = {https://doi.org/10.1145/1015706.1015772},
doi = {10.1145/1015706.1015772},
journal = {ACM Trans. Graph.},
month = aug,
pages = {630–634},
numpages = {5}
}

@misc{Han2025,
title={ARAP-GS: Drag-driven As-Rigid-As-Possible 3D Gaussian Splatting Editing with Diffusion Prior}, 
author={Xiao Han and Runze Tian and Yifei Tong and Fenggen Yu and Dingyao Liu and Yan Zhang},
year={2025},
NOeprint={2504.12788},
archivePrefix={arXiv},
primaryClass={cs.GR},
NOurl={https://arxiv.org/abs/2504.12788}, 
}

@article{Oehri2025,
author =       {Annika Oehri and Philipp Herholz and Olga Sorkine-Hornung}, 
title =        {Higher Order Continuity for Smooth As-Rigid-As-Possible Shape Modeling},
year =         {2025},
month =        {June},
day =          {6},
journal =      {Journal of Computer Graphics Techniques (JCGT)},
volume =       {14},
number =       {1},
pages =        {198--215},
NOurl =          {http://jcgt.org/published/0014/01/10/},
issn =         {2331-7418}
}

@InProceedings{Park2019,
author = {Park, Jeong Joon and Florence, Peter and Straub, Julian and Newcombe, Richard and Lovegrove, Steven},
title = {DeepSDF: Learning Continuous Signed Distance Functions for Shape Representation},
booktitle = {Proceedings of the IEEE/CVF Conference on Computer Vision and Pattern Recognition (CVPR)},
month = {June},
year = {2019}
}

@InProceedings{Mescheder2019,
author = {Mescheder, Lars and Oechsle, Michael and Niemeyer, Michael and Nowozin, Sebastian and Geiger, Andreas},
title = {Occupancy Networks: Learning 3D Reconstruction in Function Space},
booktitle = {Proceedings of the IEEE/CVF Conference on Computer Vision and Pattern Recognition (CVPR)},
month = {June},
year = {2019}
}

@inproceedings{Sitzmann2020,
 author = {Sitzmann, Vincent and Martel, Julien and Bergman, Alexander and Lindell, David and Wetzstein, Gordon},
 booktitle = {Advances in Neural Information Processing Systems},
 editor = {H. Larochelle and M. Ranzato and R. Hadsell and M.F. Balcan and H. Lin},
 pages = {7462--7473},
 publisher = {Curran Associates, Inc.},
 title = {Implicit Neural Representations with Periodic Activation Functions},
 NOurl = {https://proceedings.neurips.cc/paper_files/paper/2020/file/53c04118df112c13a8c34b38343b9c10-Paper.pdf},
 volume = {33},
 year = {2020}
}

@inproceedings{Wang2021,
 author = {Wang, Peng and Liu, Lingjie and Liu, Yuan and Theobalt, Christian and Komura, Taku and Wang, Wenping},
 booktitle = {Advances in Neural Information Processing Systems},
 editor = {M. Ranzato and A. Beygelzimer and Y. Dauphin and P.S. Liang and J. Wortman Vaughan},
 pages = {27171--27183},
 publisher = {Curran Associates, Inc.},
 title = {NeuS: Learning Neural Implicit Surfaces by Volume Rendering for Multi-view Reconstruction},
 NOurl = {https://proceedings.neurips.cc/paper_files/paper/2021/file/e41e164f7485ec4a28741a2d0ea41c74-Paper.pdf},
 volume = {34},
 year = {2021}
}

@article{Muller2022,
author = {M\"{u}ller, Thomas and Evans, Alex and Schied, Christoph and Keller, Alexander},
title = {Instant neural graphics primitives with a multiresolution hash encoding},
year = {2022},
issue_date = {July 2022},
publisher = {Association for Computing Machinery},
address = {New York, NY, USA},
volume = {41},
number = {4},
issn = {0730-0301},
NOurl = {https://doi.org/10.1145/3528223.3530127},
doi = {10.1145/3528223.3530127},
journal = {ACM Trans. Graph.},
month = jul,
articleno = {102},
numpages = {15}
}

@inproceedings{Tang2022,
 author = {Tang, Jiapeng and Markhasin, Lev and Wang, Bi and Thies, Justus and Niessner, Matthias},
 booktitle = {Advances in Neural Information Processing Systems},
 editor = {S. Koyejo and S. Mohamed and A. Agarwal and D. Belgrave and K. Cho and A. Oh},
 pages = {17117--17132},
 publisher = {Curran Associates, Inc.},
 title = {Neural Shape Deformation Priors},
 NOurl = {https://proceedings.neurips.cc/paper_files/paper/2022/file/6d09ef61aeb76be676b358f6f87b3484-Paper-Conference.pdf},
 volume = {35},
 year = {2022}
}

@article{Aigerman2022,
author = {Aigerman, Noam and Gupta, Kunal and Kim, Vladimir G. and Chaudhuri, Siddhartha and Saito, Jun and Groueix, Thibault},
title = {Neural jacobian fields: learning intrinsic mappings of arbitrary meshes},
year = {2022},
issue_date = {July 2022},
publisher = {Association for Computing Machinery},
address = {New York, NY, USA},
volume = {41},
number = {4},
issn = {0730-0301},
NOurl = {https://doi.org/10.1145/3528223.3530141},
doi = {10.1145/3528223.3530141},
journal = {ACM Trans. Graph.},
month = jul,
articleno = {109},
numpages = {17}
}

@InProceedings{Yifan2020,
author = {Yifan, Wang and Aigerman, Noam and Kim, Vladimir G. and Chaudhuri, Siddhartha and Sorkine-Hornung, Olga},
title = {Neural Cages for Detail-Preserving 3D Deformations},
booktitle = {Proceedings of the IEEE/CVF Conference on Computer Vision and Pattern Recognition (CVPR)},
month = {June},
year = {2020}
}

@InProceedings{Deng2021,
    author    = {Deng, Yu and Yang, Jiaolong and Tong, Xin},
    title     = {Deformed Implicit Field: Modeling 3D Shapes With Learned Dense Correspondence},
    booktitle = {Proceedings of the IEEE/CVF Conference on Computer Vision and Pattern Recognition (CVPR)},
    month     = {June},
    year      = {2021},
    pages     = {10286-10296}
}

@InProceedings{Zheng2021,
    author    = {Zheng, Zerong and Yu, Tao and Dai, Qionghai and Liu, Yebin},
    title     = {Deep Implicit Templates for 3D Shape Representation},
    booktitle = {Proceedings of the IEEE/CVF Conference on Computer Vision and Pattern Recognition (CVPR)},
    month     = {June},
    year      = {2021},
    pages     = {1429-1439}
}

@InProceedings{Wang2019,
author = {Wang, Weiyue and Ceylan, Duygu and Mech, Radomir and Neumann, Ulrich},
title = {3DN: 3D Deformation Network},
booktitle = {Proceedings of the IEEE/CVF Conference on Computer Vision and Pattern Recognition (CVPR)},
month = {June},
year = {2019}
}

@InProceedings{Hao2020,
author = {Hao, Zekun and Averbuch-Elor, Hadar and Snavely, Noah and Belongie, Serge},
title = {DualSDF: Semantic Shape Manipulation Using a Two-Level Representation},
booktitle = {Proceedings of the IEEE/CVF Conference on Computer Vision and Pattern Recognition (CVPR)},
month = {June},
year = {2020}
}

@InProceedings{Behrmann2019,
  title = 	 {Invertible Residual Networks},
  author =       {Behrmann, Jens and Grathwohl, Will and Chen, Ricky T. Q. and Duvenaud, David and Jacobsen, Joern-Henrik},
  booktitle = 	 {Proceedings of the 36th International Conference on Machine Learning},
  pages = 	 {573--582},
  year = 	 {2019},
  editor = 	 {Chaudhuri, Kamalika and Salakhutdinov, Ruslan},
  volume = 	 {97},
  series = 	 {Proceedings of Machine Learning Research},
  month = 	 {09--15 Jun},
  publisher =    {PMLR},
  NOurl = 	 {https://proceedings.mlr.press/v97/behrmann19a.html},
}

@InProceedings{Zhang2023,
    author    = {Zhang, Baowen and Li, Jiahe and Deng, Xiaoming and Zhang, Yinda and Ma, Cuixia and Wang, Hongan},
    title     = {Self-supervised Learning of Implicit Shape Representation with Dense Correspondence for Deformable Objects},
    booktitle = {Proceedings of the IEEE/CVF International Conference on Computer Vision (ICCV)},
    month     = {October},
    year      = {2023},
    pages     = {14268-14278}
}

@article{Smith2015,
author = {Smith, Jason and Schaefer, Scott},
title = {Bijective Parameterization with Free Boundaries},
year = {2015},
month = {7},
journal = {ACM Trans. Graph.},
volume = {34},
number = {4},
publisher = {Association for Computing Machinery},
address = {New York, NY, USA},
issn = {0730-0301},
doi = {10.1145/2766947},
articleno = {70},
nourl = {https://doi.org/10.1145/2766947},
numpages = {9}
}

@inproceedings{Tarini2004,
author = {Tarini, Marco and Hormann, Kai and Cignoni, Paolo and Montani, Claudio},
title = {PolyCube-Maps},
year = {2004},
isbn = {9781450378239},
publisher = {Association for Computing Machinery},
address = {New York, NY, USA},
NOurl = {https://doi.org/10.1145/1186562.1015810},
doi = {10.1145/1186562.1015810},
booktitle = {ACM SIGGRAPH 2004 Papers},
pages = {853–860},
numpages = {8},
location = {Los Angeles, California},
series = {SIGGRAPH '04}
}

@article{Huang2014,
author = {Huang, Jin and Jiang, Tengfei and Shi, Zeyun and Tong, Yiying and Bao, Hujun and Desbrun, Mathieu},
title = {$\ell_1$-Based Construction of Polycube Maps from Complex Shapes},
year = {2014},
issue_date = {May 2014},
publisher = {Association for Computing Machinery},
address = {New York, NY, USA},
volume = {33},
number = {3},
issn = {0730-0301},
NOurl = {https://doi.org/10.1145/2602141},
doi = {10.1145/2602141},
journal = {ACM Trans. Graph.},
month = jun,
articleno = {25},
numpages = {11}
}

@InProceedings{Chetan2025,
    author    = {Chetan, Aditya and Yang, Guandao and Wang, Zichen and Marschner, Steve and Hariharan, Bharath},
    title     = {Accurate Differential Operators for Hybrid Neural Fields},
    booktitle = {Proceedings of the IEEE/CVF Conference on Computer Vision and Pattern Recognition (CVPR)},
    month     = {June},
    year      = {2025},
    pages     = {530-539}
}

@ARTICLE{Wu2024,
  author={Wu, Kang and Zhao, Zheng-Yu and Zhang, Zheng and Liu, Ligang and Fu, Xiao-Ming},
  journal={IEEE Transactions on Visualization and Computer Graphics}, 
  title={Piecewise Developable Modeling via Implicit Neural Deformation and Feature-Guided Cutting}, 
  year={2024},
  volume={30},
  number={9},
  pages={5993-6004},
  doi={10.1109/TVCG.2023.3319487}}

@Manual{Blender,
   title = {Blender - a 3D modelling and rendering package},
   author = {{Blender Online Community}},
   organization = {Blender Foundation},
   address = {Blender Institute, Amsterdam},
   year = {2025},
   url = {http://www.blender.org},
}

@article{Finnendahl2026,
author = {Finnendahl, Ugo and Alexa, Marc},
title = {On Bending in the As-Rigid-As-Possible Deformation Energy},
journal = {Computer Graphics Forum},
volume = {45},
number = {5},
year = {2026},
  doi={10.1111/cgf.70520}}
